\documentstyle[preprint,aps]{revtex}
\def\d{{\rm d}}
\begin{document}
\draft
\title{Reduced Persistence Length and Fluctuation-Induced Interactions
of Directed Semiflexible Polymers on Fluctuating surfaces}
\author{Ramin Golestanian}
\address{Institute for Advanced Studies in Basic Sciences,
Zanjan   45195-159, Iran}
\date{\today}
\maketitle
\begin{abstract}
We consider directed semiflexible polymers embedded in a fluctuating 
surface which is governed by either surface tension or bending rigidity.
The attractive interactions induced by the fluctuations of the surface
reduce the rigidity of the polymers.  In particular, it is shown that 
for arbitrarily stiff parallel polymers, there is a characteristic separation 
below which they prefer to bend rather than stay linear.
The out-of plane fluctuations of the polymer, screen out the long-range
fluctuation-induced forces, resulting in only a short-ranged effective
attraction.
\end{abstract}

Fluid membranes with linear inclusions have shown a variety of interesting
phenomenon. These inclusions can be incorporated by a 
polymrization of membranes containing unsaturated amphiphiles
using UV irradiation \cite{Rin}. Also, some (partially) hydrophobic 
polymers that are
water soluble can become immersed or multiply anchored to
the interior of a bilayer due to hydrophobic attractions \cite{Rin}.
Thermal fluctuations, or the elastic structure of the membrane, 
in conjunction with the constraint that inclusions are attached to 
the surface, affect structural properties of the polymers. 
For example, it has been shown that\cite{Koz}, depending on the 
value of rigidity parameters, elastic properties of the membrane
may stabilize or destabilze the straight configuration of semiflexible 
polymers. 
Thermal fluctuations have been shown to induce an effective nematic-like 
orientational interaction between different segments of a polymer
\cite{Podgornik}. Polymers in turn, restricts
membrane fluctuations and alter its behavior. It was found that
cross-linked polymerization could promote wrinkling of the membranes
\cite{Dvol}, while linear polymerization causes the membrane to 
bulge and bud \cite{Sack}. 

The long distance orientational interaction between 
rod-like inclusions due to thermal surface fluctuations were 
examined in Ref. \cite{GGK}. There, it was assumed that the rods
are much more rigid than the ambient membrane. In the limit that
the separation $R$ between rods is much greater than their length $L$,
the interaction falls off as $k_BT(L/R)^4$; similar to the attractive
interaction between disk-like inclusions \cite{Goul}. It was
also shown that the interactions have the non-trivial orientational
dependence of squared dipolar interactions for inclusions on a film, 
and squared quadrupolar interactions on a membrane.
For parallel rods at close separation, $R\ll L$, the behavior is
similar to the Casimir interaction between parallel plates\cite{LiK},
resulting in an interaction proportional to $k_BT (L/R)$.

If the distances between the external objects, as well as their 
characteristic sizes, are larger than the membrane thickness, 
one can forget about the microscopic details of the membrane. In this
limit, the membrane is well-described by the elastic Hamiltonian \cite{CH},
\begin{equation}
{\cal H}_{CH}=\int \d S \left[\sigma+\frac{\kappa}{2} H^2+\bar\kappa
K \right],\label{CHH}
\end{equation}
where $\d S$ is the surface area element, and $H$, $K$ are the mean and
Gaussian curvatures respectively. The elastic properties of the surface
are described by the tension $\sigma$, and the bending rigidities
$\kappa$ and $\bar\kappa$. A finite surface tension is the most
important coupling in Eq.(\ref{CHH}) and dominates the bending terms at
long wavelengths. This is the case for films on a frame, interfaces at
short distances, and possibly closed membranes in the presence of osmotic
pressure  differences between their interior and exterior. On the other
hand,
for closed bilayers in the absense of osmotic stress, as well as for
microemulsions, the surface tension is effectively zero \cite{DGT,BL,DL}.
In these cases, the energy cost of fluctuations is controlled by the
rigidity
terms. For simplicity we shall refer to surface tension dominated surfaces
as films, and to rigidity controlled ones as membranes as in Ref. \cite{GGK}.

In this paper we examine the effect of surface fluctuations of
adsorbed {\it directed semiflexible polymers}. The directed 
semiflexible polymer is 
represented by a two dimensional position vector $\vec{r}(t)$ that indicates 
the transverse position of the polymer at point $t$ along its backbone; its 
elastic energy is then described by the Hamiltonian \cite{Nelson}
\begin{eqnarray}
{\cal H}_{DP}&=&\frac{\kappa_p}{2} \int \d t \left(\frac{\d^2 \vec{r}(t)}
	{\d t^2} \right)^2\label{DP} \\
&=&\frac{\kappa_p}{2} \int \frac{\d q}{2\pi} q^4
\left(\mid r_{\parallel,1}(q) \mid^2 
+\mid r_{\perp,1}(q) \mid^2 \right) .
\nonumber
\end{eqnarray}
Such a polymer is rigid only at distances less than a bare persistence
length $\ell_p^0\sim\kappa_p/(k_BT)$. Upon adsorption on a fluctuating
surface, the induced interactions soften the rigidity of the polymer. While
the reduction in persistence length $\ell_p$ is not appreciable 
for a polymers adsorbed on a film, there is a logarithmic reduction 
(see below) upon adsorption on a membrane. 
The softening is more dramatic for two parallel rods, which due to their 
mutual attraction want to bend towards each other. This leads to an 
instability in
the modes describing their relative {\it in-plane} fluctuations.
(Modes where the polymers fluctuate in parallel are stable).
The stiffness of the polymer is able to prevent such instabilities only
if the length of the polymer is less than a characteristic size
\begin{equation}
L_p(R) \sim \ell_{p}^{0} \left(\frac{R}{\ell_{p}^{0}} \right)^{3/4},
\label{Persist}
\end{equation}
which is much lower than $\ell_{p}^{0}$.

The {\it out of plane} fluctuations of the semiflexible  polymers have a 
dramatic effect on their ``Casimir" attraction. Consider two polymers 
adsorbed on the surface which are, on average, parallel to each other 
at separations $a \ll R < L$, where $a$ is a microscopic length scale 
such as membrane thichness. The membrane and polymers have 
thermal fluctuations but are constrained to remain attached at all times. 
In the absence of {\it out of plane} polymer fluctuations, there is
a fluctuation-induced interaction that falls off as $k_BT(L/R)$. The
{\it out of plane} fluctuations actually screen out this interaction,
resulting in a  {\it short-ranged} attraction. The precise form of the
interaction (presented in Eq.(\ref{Fperp}) below) is not particularly 
illuminating. We have instead approximately fitted the variations of 
the free energy to a form 
\begin{equation}
\frac{F^{\perp}(R)}{L}=-\frac{A \;k_{B}T}{2 \pi R} \exp\left[-b 
		\left(\frac{R}{\lambda}\right)^{\alpha}\right],\label{Perp}
\end{equation}
where $A_F=\pi^2/12$, $A_M=2.90514$, $b_F=3.32$, $b_M=4.17$,
$\alpha_F=0.679$ and $\alpha_M=0.397$ are numerical constants 
corresponding to the cases of films and membranes respectively.
The ``screening'' lengths that determine the range of the interactions
are defined as $\lambda_F=(\kappa_p/\sigma)^{1/3}$
for films and $\lambda_M=\kappa_p/\kappa$ for membranes.

To perform the calculations we start with a thermally fluctuating planar
membrane subject to the Hamiltonian in Eq.(\ref{CHH}). For the membrane
case, we assume that its size is well below the persistence length
$\xi_p$\cite{DGT}. In this limit, the membrane experiences only small
fluctuations about a flat state. We may then parametrize the membrane 
surface with a height function $h(r)$, and approximate the full 
coordinate-invariant Hamiltonian in Eq.(\ref{CHH}) by the Gaussian form
${\cal H}_M=\kappa/2 \int\d^2r \left(\nabla^{2}h(r)\right)^2$.
For the case of films, the corresponding Gaussian approximation
can be written as 
${\cal H}_F=\sigma/2\, \int\d^2r\left(\nabla h(r)\right)^2$,
which is valid in the limit where $\sigma a^2/k_{B}T \gg 1$,
where $a$ is a microscopic length.
We also parametrize the polymers as ${\bf R}_{\alpha}(t)=
({\bf r}_{\alpha}(t),r_{\perp,\alpha}(t))$ where the ``perpendicular''
axis is the same as the height function, while ${\bf r}$ stands for 
the in plane coordiantes. In calculating fluctuation induced interactions,
we assume that the relaxation of surface modes takes place much
faster than that of the polymers. We thus assume fixed (quenched)
configurations for the polymers, and integrate over the constrained
surface fluctuations, following Ref. \cite{LiK}, as
\begin{eqnarray}
\exp(-\beta {\cal H}_{eff})&=&\frac{1}{{\cal Z}_{0}}\int {\cal D}h({\bf r})
	\prod_{\alpha=1}^{2} \delta \left\{h({\bf r}_{\alpha}(t_{\alpha}))
	-r_{\perp,\alpha}(t_{\alpha}) \right\} {\rm e}^{-\beta {\cal H}}
	\label{Heff1} \\
&=&\frac{1}{{\cal Z}_{0}}\int {\cal D}h({\bf r}) \prod_{\alpha=1}^{2}
\int {\cal D}\psi_{\alpha}(t_{\alpha}) \exp\left\{-\beta {\cal H}+i 
\sum_{\alpha=1}^{2}\int \d t_{\alpha} \psi_{\alpha}(t_{\alpha})
\left[h({\bf r}_{\alpha}(t_{\alpha}))-r_{\perp,\alpha}(t_{\alpha}) \right]
	\right\} \nonumber \\
&=&\int \prod_{\alpha=1}^{2} {\cal D}\psi_{\alpha}(t_{\alpha})
	{\rm e}^{-{\cal H}_1}, \nonumber 
\end{eqnarray}
where $\cal H$ can be either ${\cal H}_F$ or ${\cal H}_M$, and
\begin{eqnarray}
{\cal H}_1&\equiv&\frac{1}{2} \Psi^{T} {\bf M} \Psi+ i \Psi^{T} r_{\perp} 
\label{H1} \\
&=&\frac{1}{2}\sum_{\alpha,\beta=1}^{2} \int \d t_{\alpha}            
\d t_{\beta} \psi_{\alpha}(t_{\alpha})
G\left({\bf r}_{\alpha}(t_{\alpha})-{\bf r}_{\beta}(t_{\beta}) \right)
\psi_{\beta}(t_{\beta}) +i \sum_{\alpha=1}^{2} \int \d t_{\alpha}  
\psi_{\alpha}(t_{\alpha})  r_{\perp,\alpha}(t_{\alpha}), \nonumber
\end{eqnarray}
and the expressions for $G({\bf r}-{\bf r}')$ read 
\begin{eqnarray}
G_F({\bf r}-{\bf r}')&=&\frac{k_{B}T}{\sigma} \left(\frac{1}{-\nabla^2}
\right)_{{\bf r}{\bf r}'}, \label{GF} \\
G_M({\bf r}-{\bf r}')&=&\frac{k_{B}T}{\kappa} \left(\frac{1}{\nabla^4}
\right)_{{\bf r}{\bf r}'}, \label{GM}
\end{eqnarray}
corresponding to films and membranes respectively. Hence, the resulting
expression reads
\begin{equation}
\beta {\cal H}_{eff}=\frac{1}{2} \ln \det \left\{ {\bf M}
	[{\bf r}_{\alpha}(t_{\alpha})]\right\}+\frac{1}{2}
	\sum_{\alpha,\beta=1}^{2} \int \d t_{\alpha} \d t_{\beta}           
	\;r_{\perp,\alpha}(t_{\alpha}) {\bf M}^{-1}(t_{\alpha},t_{\beta})
	r_{\perp,\beta}(t_{\beta}). \label{Heff2}
\end{equation}
To calculate $\ln \det\{{\bf M}\}$ we parametrize the in-plane coordinates
of the polymers as ${\bf r}_{1}(t)=(t,r_{\parallel,1}(t))$ and
${\bf r}_{2}(t)=(t,R+r_{\parallel,2}(t))$, and follow closely the method of
Ref. \cite{LiK}. We obtain
\begin{eqnarray}
\beta {\cal H}_{eff}&=&\frac{L}{2}\int \frac{\d q}{2\pi} \;\ln(N(q))  
		-\frac{1}{2}\int \frac{\d q}{2\pi} \;A(q)\;
		\left(\mid r_{\parallel,1}(q) \mid^2
		+\mid r_{\parallel,2}(q) \mid^2 \right) 
		\label{Heff3} \\      
		&+&\frac{1}{2}\int \frac{\d q}{2\pi} \;B(q)\;
		\left( r_{\parallel,1}(q) r_{\parallel,2}(-q)
		+r_{\parallel,1}(-q) r_{\parallel,2}(q) \right)
		\nonumber \\
		&+&\frac{1}{2}\int \frac{\d q}{2\pi} \;
		\frac{G(q,0)}{N(q)}\;
		\left(\mid r_{\perp,1}(q) \mid^2
		+\mid r_{\perp,2}(q) \mid^2 \right) 
		\nonumber \\      
		&-&\frac{1}{2}\int \frac{\d q}{2\pi} \;
		\frac{G(q,R)}{N(q)}\;
		\left( r_{\perp,1}(q) r_{\perp,2}(-q)
		+r_{\perp,1}(-q) r_{\perp,2}(q) \right)
		+O(r^3), \nonumber
\end{eqnarray}                
where the kernels are defined as
\begin{eqnarray}
A(q)&=&\int \frac{\d p}{2\pi} \; \frac{G(p,0)}{N(p)}\;
\left.\frac{\partial^2 G(p+q,R)}{\partial R^2} \right|_{R=0} \label{A(q)}\\
&+&\int \frac{\d p}{2\pi} \;
\left(\frac{\partial G(p,R)}{\partial R}\right)^2   \;
\left(\frac{G(p,0) G(p+q,0)+G(p,R) G(p+q,R)}{N(p)N(p+q)} \right),
	\nonumber
\end{eqnarray}
and
\begin{eqnarray}
B(q)&=&\int \frac{\d p}{2\pi} \; \frac{G(p,R)}{N(p)}\;
\frac{\partial^2 G(p+q,R)}{\partial R^2}  \label{B(q)}\\
&+&\int \frac{\d p}{2\pi} \;
\frac{\partial G(p,R)}{\partial R} \;
\frac{\partial G(p+q,R)}{\partial R} \;
\left(\frac{G(p,0) G(p+q,0)+G(p,R) G(p+q,R)}{N(p)N(p+q)} \right),
	\nonumber
\end{eqnarray}
with
\begin{eqnarray}
G(p,R)&=&\int \d t \;{\rm e}^{ipt} G(t,R), \label{Gdef} \\
N(p)&=&\left(G(p,0)\right)^2 - \left(G(p,R)\right)^2. \nonumber
\end{eqnarray}

The Fourier transformed Green's functions for  films and                 
membranes are respectively
\begin{eqnarray}
G_F(p,R)&=&\frac{k_{B}T}{\sigma} \;\frac{{\rm e}^{-\mid p \mid R}}
{2\mid p \mid} \label{GFGM} \\
G_M(p,R)&=&\frac{k_{B}T}{\kappa} \;\frac{1}{4\mid p \mid^3} 
\left(1+\mid p \mid R \right) {\rm e}^{-\mid p \mid R}. \nonumber 
\end{eqnarray}

On the slower time scales after surface modes have come to equilibrium,
fluctuations of the semiflexible polymers are governed by the Hamiltonian
${\cal H}_{tot}={\cal H}_{DP}+{\cal H}_{eff}$ (see Eqs.(\ref{DP})
and (\ref{Heff3})). In the linearized theory, the modes parallel and
perpendicular to the surface are independent.
We shall first describe the behavior of a single polymer, obtained
from the limit $R\to\infty$, with $B(q)=G(q,R)=0$.
The {\it out of plane} modes of the semiflexible polymer turn out to 
be stable.

More care is necessary for the {\it in plane} fluctuations.
Close examination of the kernels $A(q)$ and $B(q)$ shows that
they have a very well-behaved expansion in powers of $qR$. So
in the long wavelength limit, we can
keep only the first few terms in the expansion. As $R\to\infty$, i.e.
for a single polymer, $B_F(q)=B_M(q)=0$,
$A_F(q)=(1/6\pi) q^3$ and $A_M(q)=(1/\pi)[\ln(L/a)-1] q^3$. In this 
case, the in-plane modes have eigenvalues equal to
$(\kappa_p/k_{B}T) q^4-(1/6\pi) q^3$ for the case of films and
$(\kappa_p/k_{B}T) q^4-(1/\pi)[\ln(L/a)-1] q^3$ for membranes.
To prevent instabilty, all eigenvalues must remain positive. 
This places restrictions on the length of the polymer, leading to new
persistence lengths. One can easily see that for films the persistence
length doesn't differ from that of the non-embedded polymer namely,
$\ell_{p}^{F}\sim\ell_{p}^{0}$. However, for the
membrane case the persistence length is logarithmically decreased
namely, $\ell_{p}^{M}[\ln(\ell_{p}^{M}/a)-1] \sim \ell_{p}^{0}$.

For the case of two polymers (or finite $R$) one finds 
$A(q)=B(q)=(A/\pi)/R^3$ to the leading order, where $A$ is defined 
as in Eq.(\ref{Perp})
for films and membranes (the $R$ independent terms already considered
above are neglected here). The normal modes of in-plane oscillations
describe either the parallel motion of the polymers with eigenvalue
$(\kappa_p/k_{B}T) q^4-A(q)+B(q)$,
or their relative motion with eigenvalue
$(\kappa_p/k_{B}T) q^4-A(q)-B(q)$ (see Eq.(\ref{Heff3})).
One can see that to the leading order, the parallel modes are 
stable (and this indeed holds to any order), while  the relative 
distance modes may be unstable.
This new instability introduces a new characteristic length
for the combined system of two semiflexible polymers adsorbed on a fluctuating
surface, defined in Eq.(\ref{Persist}). As one can immediately see,
this $L_p$ is considerably smaller than the persistence length of a single 
polymer. Since it is a function of the separation of the polymers,
one can estimate the closest separation that two rigid rods can have,
in order not to completely lose their rigidity. Assume that the rods are
very rigid while they are free ($L\ll \ell_{p}^{0}$), then the critical
``rigidity-blurring'' separation occurs when $L\sim L_p$ which reads
$R_c \sim L(L/\ell_{p}^{0})^{1/3}$. For example for relatively rigid
rods with $L/\ell_{p}^{0} \sim 10^{-3}$, the critical separation is 
only about $R_c/L \sim 1/10$. Since
the unstable modes are the relative distance modes, this result
give us the following picture. The rods move towards each other 
due to the attractions and when the get closer than $R_c$, they
start to bulge out. If there exist some short distance
anchoring interaction between the polymers, they would bend
towards each other to form bound states (like a necklace). However,
if short distance interactions are repulsive, they would bounce
back against each other and stretch out from either side.

In the absence of polymer fluctuations $r_\parallel=r_\perp=0$, the
interaction between rods in given by the first term in Eq.(\ref{Heff3})
and decays as $L/R$. Assuming that the lengths $L$ and $R$ are selected
such that the stability conditions given above are satisfied, a new
net fluctuation-induced interaction between {\it semiflexible} polymers 
is obtained by integrating over both
$r_\parallel$ and $r_\perp$ in Eq.(\ref{Heff3}), keeping the
center of mass distance fixed at $R$. However, integration
over $r_\parallel$ in this case will depend on the low momentum
cut-off and we don't consider that contribution here, since it doesn't
have any particular effect, other than introducing some small corrections
to the long range ``Casimir'' interaction. 
The contribution of out of plane modes to the resulting constrained 
free energy is given by
\begin{equation}
\frac{F^{\perp}}{L}(R)=\frac{k_{B}T}{2} \int \frac{\d q}{2\pi} \;
\ln \left[\left(\frac{\kappa_p}{k_{B}T}\right)^2 q^8
+2 \frac{\kappa_p}{k_{B}T} q^4 \frac{G(q,0)}{N(q)}+\frac{1}{N(q)}
\right]. \label{Fperp} 
\end{equation}
It is interesting to note that the above expression cancels the
the long-range ``Casimir'' interaction from  $\ln N(q)$ in 
Eq.(\ref{Heff3}), and the remaining expression do not result in a
long-ranged interaction anymore. 
So as soon as the rods have a {\it finite} rigidity, no matter
how large, and they are allowed to have out-of plane fluctuations,
the interaction will be ``screened''.
However, the ``screening length'' is defined
by the rigidity and for separations that are much smaller than
this ``screening length'', one may neglect this effect. 

Plugging in the explicit expressions for the Green's functions 
from Eqs.(\ref{GFGM}) and (\ref{Gdef}), we obtain $F^{\perp}_F$ 
and $F^{\perp}_M$ in terms of integrals that do not have analytic 
expressions. Since those forms are not very illuminating, we 
give the best fit for the result in terms of exponentials of power laws. 
These result are quoted in Eq.(\ref{Perp}) for both films and membranes
and Fig.~1 compare the fits and the interactions. 
It is important to note that the integral expressions are non-analytical
at $R=0$ as it reveals when one tries to make a taylor expansion. That's
why fitting is done with non-analytical powers. This is also checked
against the stability of the fit.

In conclusion, we have shown that the rigidity of semiflexible polymers
is lowered upon adsorption to a fluctuating surface. The effect is enhanced
when two such polymers are brought in close proximity. The out of plane
fluctuations of the polymer, in turn, soften the boundary conditions that
give rise to long-ranged ``Casimir" attraction. After they are integrated out, 
the attraction between two parallel polymers becomes short-ranged.

I would like to acknowledge stimulating discussions with M. Kardar, 
M.R.H. Khajehpour and M. Goulian. 
I also wish to thank M. Kardar for careful reading of the
manuscript and giving critical comments.
The work is supported by the Institute for Advanced Studies
in Basic Sciences, Gava Zang, Zanjan, Iran.

\begin{figure}
\caption{Dimensionless energy $f=F^{\perp}/F_{C}$ plotted versus the 
dimensionless separation $x=R/\lambda$, where $F_{C}$ is the long-ranged 
Casimir interaction. The solid curves represent the best fit to the
exact data point obtained from numerical integration.}
\end{figure}

\end{document}